\documentclass[12pt,fleqn,twoside]{article}
\usepackage{espcrc1}

\usepackage{graphicx}
\input epsf

\newcommand{\AmS}{{\protect\the\textfont2
  A\kern-.1667em\lower.5ex\hbox{M}\kern-.125emS}}
\def\0{\over } \def\2{{1\over2}} \def\4{{1\over4}}
\def\5{\hat } \def\6{\partial }

\def\({\left(} \def\){\right)} \def\<{\langle } \def\>{\rangle }

\def\simge{\mathrel{%
   \rlap{\raise 0.511ex \hbox{$>$}}{\lower 0.511ex \hbox{$\sim$}}}}
\def\simle{\mathrel{
   \rlap{\raise 0.511ex \hbox{$<$}}{\lower 0.511ex \hbox{$\sim$}}}}

\newcommand{\bea}{\begin{eqnarray}}
\newcommand{\eea}{\end{eqnarray}}
\newcommand{\be}{\begin{equation}}
\newcommand{\ee}{\end{equation}}

\newcommand \beq{\begin{eqnarray}}
\newcommand \eeq{\end{eqnarray}}

\def\del{\partial }  

\title{The Entropy of the Quark-Gluon Plasma}

\author{J.-P. Blaizot\address{Service de Physique Th\'eorique,
CEA Saclay\\
91191 Gif-sur-Yvette cedex, France
}}
       
\begin{document}
\maketitle
\begin{abstract}
The entropy of the quark-gluon plasma can be calculated from QCD using
(approximately) self-consistent approximations. Lattice results for
pure gauge theories are accurately reproduced down to temperatures
of the order of 2.5$T_c$. Comparisons with other approaches to the
thermodynamics of the quark-gluon plasma are briefly discussed.\vspace{1pc}
\end{abstract}


\section{Introduction}

As we were reminded, the Bielefeld
meetings which took place in 1980 and 1982 have played a decisive role in
our field. It should be added that, since then, the Bielefeld group has played a leading role
in the study of QCD thermodynamics with lattice calculations. There is one
particular result  of that time  that I wish to recall here
because it is central to the topic of my talk. This result is that of the
lattice calculation of the energy density of SU(2) pure gauge theory
\cite{Engels:1982qx}. It is reproduced in
Fig.~1 where one sees clearly the phase transition and the 
approach to the ideal gas limit at high temperature confirming expectations based on
asymptotic freedom.

Modern results for the SU(3) equation of state \cite{Boyd:1996bx} are displayed in
Fig.~2.  The talk will focus on the slow approach to the  free
gas limit of the thermodynamical functions at high $T$.  The
question  that I wish to address is whether one can understand
the deviation from the ideal gas behaviour with weak coupling
calculations. Beyond this question is, as we shall see, another
one  related to the relevance of the notion of quasiparticles
in the description of the quark-gluon plasma.

\begin{figure}\label{fig:SU2}
\epsfxsize=0.70\hsize
\hfil\epsfbox{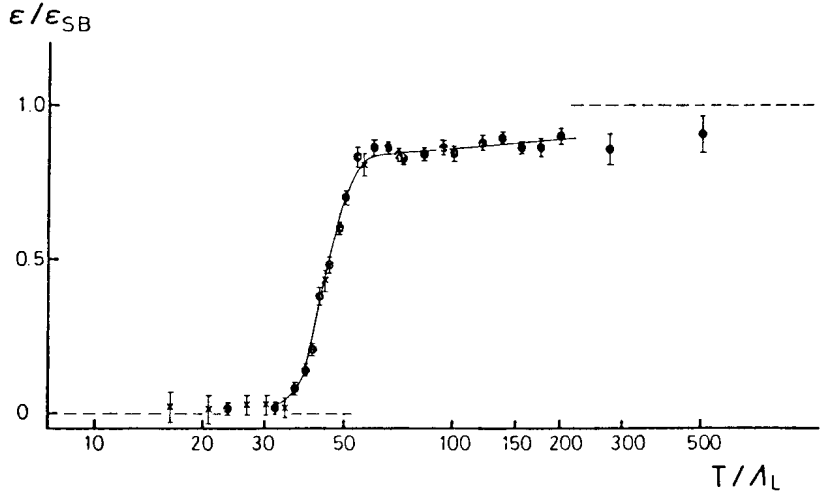}\hfil
\caption{Energy density of a SU(2) gluon plasma normalized to that of an ideal 
gas of gluons, as a function of the temperature. From Ref.~[1].}
\end{figure}

The motivation for considering  the quark-gluon plasma as a weakly coupled system is of
course asymptotic freedom, which leads us to expect that the effective coupling $g$ to be
used in thermodynamical calculations should be small if the temperature $T$ is high
enough:
\beq 
\alpha_s(\mu)\,\equiv\,{g^2\over 4\pi}\,\propto\,
      \frac{1}{\ln( \mu/\Lambda_{QCD})}\,,
\eeq
with typically $ \mu \,\simeq\,2\pi T $.
But we know that, even in cases where the coupling is small, strict
perturbation theory cannot be used. Technically, infrared divergences occur in
high order calculations, and various resummations are needed to
get meaningful results.

Some of the difficulties
 of perturbation theory are already visible on the lowest orders which, for SU(3)
gluons, read:
\beq
\frac{P}{P_0}=1-\frac{15 }{4}\left( \frac{\alpha_s
}{\pi}\right)+30
\left(\frac{\alpha_s }{\pi}\right)^{3/2},
\eeq
with $P_0$ the ideal gas pressure:
\beq
\qquad P_0=8 \,\frac{\pi^2
T^4}{45}.
\eeq
 The large coefficient of the
$g^3$ term  makes its contribution to the pressure   comparable to that of the
order $g^2$ when $g\simge 0.8$, or $\alpha_s\simge 0.05$; larger coupling makes the pressure larger than that of the ideal gas. Now it is worth
emphasizing that the term of order
$g^3$ emerges from an infinite resummation (strict perturbation theory at finite order
would lead to a polynomial in $g^2$); it is the leading term of the resummed expression
when $g$ is small. 
 We shall later argue that the underlying strategy (that of performing an infinite
resummation which is then re-expanded in powers of $g$) may not be used to extrapolate to
large coupling.  Physically the need for resummation arises from the existence
of collective excitations in the system, whose properties are not well captured by 
 perturbation theory.

\begin{figure}\label{fig:entropieSU3}
\epsfxsize=0.70\hsize
\hfil\epsfbox{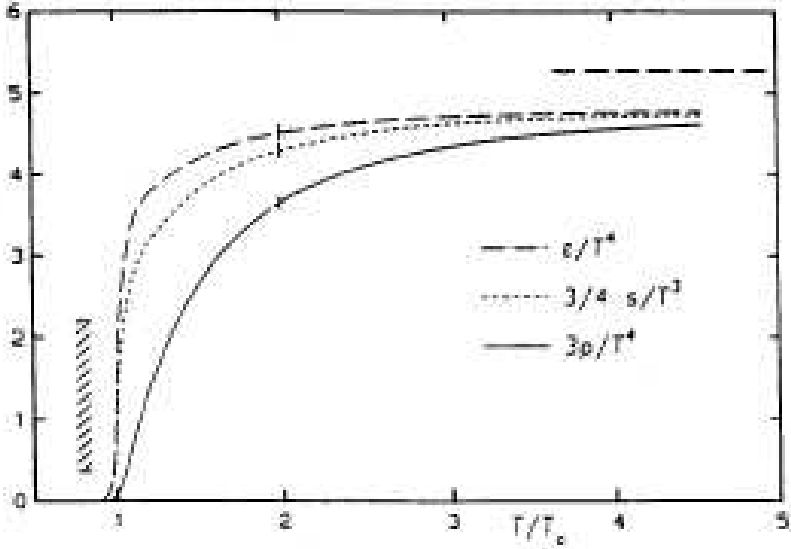}\hfil
\caption{The thermodynamical functions in SU(3) gauge theories. From Ref.~[2].}
\end{figure}

A lot of efforts have been devoted in the recent years to push perturbative calculations
of the pressure to the highest calculable
order, namely order $g^5$ \cite{QCDP,BN,Braaten01}.
The contributions of order $g^4$ and $g^5$ 
do not resolve the difficulties met at order $g^3$: the values of the pressure
obtained by adding successively these high order contributions oscillate
wildly, and reveal a strong dependence on the renormalization scale.  Attempts
have been made to construct smooth extrapolations based on  the first terms of
the series, using Pade approximants
\cite{Hatsuda:1997wf,Chiku:1998kd}  or Borel summation techniques
\cite{Parwani:2001rr,Parwani:2001am}. The resulting expressions are 
indeed  smooth functions of the coupling, better behaved than polynomial
approximations truncated at order $g^5$ or lower, with a weak dependence on the
renormalization scale.  However these techniques, which  can be powerful, offer 
little physical insight.

A more appealing physical picture is provided by quasiparticle models in which one
assumes that the dominant effect  of the interactions can be incorporated in the spectral
properties of suitably defined quasiparticles.  
 It is useful to review briefly the simplest version 
of such models, limiting ourselves here to bosonic quasiparticles
\cite{Peshier}. One assumes that the system is composed of non interacting massive quasiparticles with
energies $E_k$:  
\beq E_k^2=k^2+m^2(T) \eeq
with the mass $m$ some function of the temperature. The  entropy density is that of the ideal gas:
\beq  s=\int \frac{d^3 k}{(2\pi)^3} \left[ (1+N_k)\log (1+N_k)-N_k\log
N_k\right],\nonumber\\ 
\eeq where 
\beq
\qquad N_k=\frac{1}{{\rm e}^{E_k/T}-1}.
\eeq 
The pressure $P$ and the energy density $\epsilon$ are then given by the
corresponding ideal gas expressions, to within a function $B(T)$ adjusted so
as to satisfy thermodynamical identities. That is, one sets:
\beq  P=-T\int \frac{d^3 k}{(2\pi)^3} \log\left( 1-{\rm e}^{-E_k/T}\right) -B(T)\eeq
\beq \epsilon =\int \frac{d^3 k}{(2\pi)^3} N_k E_k +B(T). \eeq Such a parametrization
obviously fulfills the identity $\epsilon+P=Ts$.  The function $B(T)$ is
then determined by requiring that 
\beq s=\frac{dP}{dT} .\eeq
Such a model  has been shown to provide a quite good basis for a fit to lattice
results~\cite{Levai:1997yx}. Recent extensions of the model can be found in
Ref.~\cite{Schneider:2001nf}. 

 Finding
a microscopic justification for this quasiparticle picture is the main motivation of the
work that I shall present, summarizing results obtained in collaboration with E. Iancu and
A. Rebhan \cite{Blaizot:1999ip,Blaizot:1999ap,Blaizot:2000fc,RebhanHere}.  Our analysis
is based on the entropy, for which the simple formulae above already suggest why it
is simpler than the other thermodynamical functions:  in both the pressure
 and the energy density  there are interaction
contributions, which cancel out in
the entropy, leaving an expression in which all interaction effects are
summarized in the properties of quasiparticles (in the simple model, the
temperature dependent mass $m(T)$). We shall see later under which conditions
this simple picture is valid. 

At this stage, it is useful to have a more general characterization  of the effects of weak
interactions in a quark-gluon plasma. Let me then go through a short  digression
about the basic scales and degrees of freedom of such a system when the coupling is small
\cite{Blaizot:2001nr}.

\section{Scales and degrees of freedom. Hard thermal loops}

In the absence of interactions, the quark-gluon plasma is a gas of massless particles
with energy $E_k=k$. Small interactions  have little effects on most particles which have 
momentum $k\sim T$, but they strongly modify the propagation of low momentum modes.  Since
the coupling to gauge fields occurs typically through covariant derivatives,
$D_x=\del_x+igA(x)$,  the effect of interactions on particle motion depends indeed on the
momentum of the excitation and the magnitude of the gauge field. 
A measure of the strength of the gauge fields  is obtained from the magnitude of their
thermal fluctuations, that is
$\bar A\equiv \sqrt{\langle A^2(t,{\bf x})\rangle}$. In equilibrium $\langle
A^2(t,{\bf
x})\rangle$ is independent of $t$ and ${\bf x}$ and given, in the non interacting case, 
by:
\beq\label{fluctuationsA}
\langle A^2\rangle\approx
\int \frac{{\rm d}^3
k}{(2\pi)^3}\frac{N_k}{E_k}.
\eeq
Here we shall use this formula also in the interacting case,
assuming that  the  effects of the interactions can be accounted for  simply
by a  change of $E_k$ (a more complete calculation is presented in
Ref.~\cite{Blaizot:2001nr}).
  
  For
the plasma particles   $E_k=k\sim T$ and  $\langle A^2\rangle_T\sim T^2$.
The associated electric (or magnetic) field fluctuations 
$
\langle (\del A)^2\rangle_T \sim
T^4
$ are a dominant contribution to the plasma energy density.
 As already mentioned, these short wavelength,
or {\it hard}, gauge field
fluctuations produce a small perturbation on the motion of a plasma particle with
momentum $k\sim T$, since $g\bar A_T\sim gT\ll k$. However, this is not so for an
excitation at the momentum scale 
$k\sim gT$, since then  the two terms in the covariant derivative
$\del_x$ and $g\bar A_T$ become comparable. That is,  an
excitation with momentum $gT$ is non perturbatively affected 
by the hard thermal fluctuations. And indeed, the scale
$gT$ is that  at which collective phenomena develop. The emergence of the Debye
screening mass
$m_D\sim gT$ is one of the simplest examples of such phenomena. 

The fluctuations at the {\it soft} scale
$gT\ll T$ can be  described by classical
fields. In fact the associated occupation numbers
$N_k$ are large, and
accordingly one can replace
$N_k$ by $ T/E_k\sim 1/g$ in eq.~(\ref{fluctuationsA}).
Introducing  an upper cut-off $gT$ in the momentum integral, one  then
gets:
\beq\label{fluctu}
\langle A^2\rangle_{gT} \sim \int^{gT}{\rm d}^3k \, \frac{T}{k^2}\sim
gT^2.
\eeq
Thus $\bar A_{gT}\sim \sqrt{g} T$ so that $g\bar A_{gT}
\sim g^{3/2}T$ is still of higher order than the kinetic term $\del_x\sim gT$.
In that sense the soft modes with $k\sim gT$ are still perturbative, i.e. their
self-interactions can be ignored in a first approximation. Note however that
they generate contributions to physical observables which are not analytic in
$g^2$, as shown by the example of the order
$g^3$ contribution to the
pressure (or as in eq.(\ref{fluctu}) above).

At the lower momentum scale $k\sim g^2T$, 
unscreened magnetic fluctuations play a dominant role.  The contributions of such  {\it
ultrasoft } fluctuations is 
\beq\label{fluctg2t}
\langle A^2\rangle_{g^2T}\sim T\int_0^{g^2T}{\rm d}^3k \frac{1}{k^2}\,
\sim\,g^2 T^2,\eeq
so that $g\bar A_{g^2T}\sim g^2T$ is now of the same order as the 
 ultrasoft 
derivative $\del_x\sim g^2T$: the fluctuations are no longer perturbative. This
is the origin of the breakdown of perturbation theory in high temperature QCD.
A more detailed analysis reveals that 
the fluctuations at scale $g^2T$ come  from the zero Matsubara frequency and
correspond therefore to those of a three dimensional theory of static fields.  

Having identified these various scales, we note now that the thermodynamical functions at
high
temperature are dominated by hard degrees of freedom. The quasiparticle picture that we
shall develop takes into account consistently the contributions of the hard degrees of
freedom together with those of the long wavelength collective excitations.
 We shall comment
later on
 the techniques based on dimensional reduction which allows a treatment of 
the  ultrasoft contributions. 

There exists a well developed effective theory for the soft collective excitations that
we briefly outline now. 
These soft
excitations can be described in terms of
{\it average  fields} which obey classical equations of motion. In QED these equations
are Maxwell equations:
\beq\label{maxwell00}
\del_\mu F^{\mu\nu}=j_{ind}^\nu+j_{ext}^\nu
\eeq
with a  source term  composed of an external
perturbation $j_{ext}^\nu$, and an extra contribution $j_{ind}^\nu$ 
 referred to as the {\it induced current}. The induced current, which summarizes the
effects of the hard degrees of freedom,  can be obtained
using linear response theory. It is of the form:
\beq\label{responsej}
j_\mu^{ind}=\int {\rm d}^4 y\, \Pi^R_{\mu\nu}(x-y) A^\nu (y),
\eeq
where $A^\nu (y)$ is the total gauge field (including the induced potential) and the
(retarded) response function
$\Pi^R_{\mu\nu}(x-y)$ is also referred to as the polarization tensor.
In leading order in weak coupling, this
polarization tensor is given by the one-loop approximation, and can be written as
$\Pi(\omega,p)=g^2T^2f(\omega/p,p/T)$ with $f$ a dimensionless function. Furthermore, the
relevant kinematics is such that the incoming momentum is soft while the loop momentum is
hard. 
 Thus, in
leading order in
$p/T\sim g$,
$\Pi$ is of the form $g^2T^2f(\omega/p)$. This particular contribution of the
one-loop polarization tensor is an example of what has been called a ``hard
thermal loop''
\cite{BP}. 

In a non Abelian theory,
linear response is not sufficient:
constraints due to gauge symmetry force us to take into account
specific non linear effects. The relevant generalization of
the Yang-Mills equation reads \cite{QCD,BIO} :
\bea\label{ava}
D_\nu F^{\nu\mu}=\Pi_{\mu\nu}^{ab}A_b^\nu
+\frac{1}{2}\,\Gamma_{\mu\nu\rho}^{abc} A_b^\nu A_c^\rho+\,...
\eea
where the induced current in the right hand side is non-linear: when
expanded in powers of $A^\mu_a$, 
it generates an infinite series of hard thermal loops (self-energy and vertex
corrections). 

In the next section, we shall consider various reorganizations of the
perturbative expansion inspired by the previous analysis  and which try 
to accommodate the non perturbative features associated
with the various scales. 

\section{Reorganizing perturbative expansion}

As we have seen, one of the main effects of the thermal fluctuations is to generate a
mass for the soft modes. This is not peculiar to gauge theories, but occurs also in the
simpler case of a scalar field. In this case, the mass generation can be taken into
account non perturbatively  by a very simple reorganization of the perturbative
expansion, leading to the so-called   ``screened perturbation theory''
\cite{Karsch:1997gj}.  One writes:
\beq
{\cal L}&=& {\cal L}_0-\frac{1}{2}m^2\phi^2+
\frac{1}{2}m^2\phi^2+{\cal L}_{int}\nonumber\\
&=& {\cal L}_0^\prime +{\cal L}_{int}^\prime \, ,
\eeq
with ${\cal L}_0^\prime={\cal L}_0-(1/2)m^2\phi^2$. 
The mass $m\sim gT$ which appears here is for instance that obtained in leading order from
the hard thermal loop.  A perturbative expansion in
terms of screened propagators (that is keeping the screening mass as
a parameter, i.e. not as a perturbative correction to be expanded out) 
has been shown to be quite stable  with good
convergence properties. However, 
$m$ makes the propagators explicitly temperature dependent, and  the ultraviolet
divergences which occur in high order calculations become temperature
dependent.  Although we know that temperature dependent infinities should
eventually cancel out, the systematics of such cancellations is not
immediately transparent.    

In the case of gauge theory, the effect of the interactions
is more complicated than just generating a mass. But we know how to determine
the dominant corrections to the self-energies. When the momenta are soft, these
are given by the hard thermal loops discussed above. By adding these
corrections to the tree level Lagrangian, and subtracting them from the
interaction part, one generates the so-called hard thermal loop perturbation
theory \cite{Andersen:2000yj,Andersen:1999fw,Andersen:1999sf}:
\beq
{\cal L}={\cal L}_0+{\cal L}_{HTL}- {\cal L}_{HTL}+{\cal L}_{int}
= {\cal L}_0^\prime +{\cal L}_{int}^\prime\, .
\eeq
 The resulting
perturbative expansion is made complicated however by the non local nature of the hard
thermal loop action, and by the necessity of introducing temperature dependent counter
terms.  Also, in such a scheme, one is led to use the hard thermal loop approximation in 
kinematical regimes where it is not justified.

Another approach that I wish to mention at this stage is that based on dimensional
reduction. This approach puts emphasis on the non perturbative, very long wavelength
fluctuations.  It has been shown in Refs.~\cite{Braaten:1995na,Braaten:1996ju,Braaten:1996jr} how, in principle, an
effective theory could be constructed to deal with this
particular problem by marrying analytical techniques (to determine the
coefficients of the effective theory) and numerical ones (to solve the non
perturbative 3-dimensional effective theory). The resulting effective theory
is a 3-dimensional theory of static fields, with Lagrangian:
\beq
{\cal L}_{eff}=\frac{1}{4}
(F^a_{ij})^2+\frac{1}{2}(D_iA_0^a)^2+\frac{1}{2}m_D^2(A_0^a)^2+ \lambda 
(A_0^a)^4
+\delta{\cal L},
\eeq
with $
D_i=\partial_i-ig\sqrt{T}A_i
$. This strategy has been applied recently to the calculation of the free
energy of the quark-gluon plasma a high temperature \cite{Kajantie:2001iz}.
 This
technique of dimensional reduction puts a special weight on
the static sector (it singles out the contributions of the zero Matsubara frequency), and
a major effort is devoted to the calculation of the coefficients of the effective
Lagrangian (which contain the dominant contribution to the thermodynamical functions). 

In contrast, the approach that we have developed  keep the full spectral information that
one has about the plasma excitations.  In a sense, it is close to the approach based on
hard thermal loop perturbation theory. It differs in that we focus on the simplest 
of the thermodynamical function, namely the entropy, for which indeed interesting
cancellations occur which allow us to bypass some of 
the difficulties of hard thermal
loop perturbation theory.

\section{Propagator renormalisation. Skeleton expansion}

As already stated, we want to incorporate full spectral information about quasiparticles
in thermodynamical calculations, and we are therefore led to carry out a propagator
renormalisation. Although in gauge theory vertex renormalisation should be carried  along
with propagator  renormalisation, we shall see here that special kinematical
conditions allow us to develop useful approximations without having
to do so. 

The formalism that we shall rely on was developed long ago in the context of the non
relativistic many-body problem. It is based on an expression for the pressure as a
functional of full propagators, first written out by  Luttinger and Ward
\cite{Luttinger:1960}. Systematic developments using either
functional Legendre transforms or diagrammatic formalisms were presented 
later  by  De Dominicis and  Martin \cite
{Cirano64}. In implementing these methods to field theory one meets extra difficulties
related to ultraviolet divergences. I
now briefly outline the general method, writing explicitly only formulae appropriate for a
scalar field (similar ones hold for QCD). 

The free
energy  ${\cal F}$ can be written as the following functional
 of the full propagator $D$: 
\beq\label{functional}
{\cal F}[D]=\frac{1}{2}{\rm Tr}\ln D^{-1}-\frac{1}{2}{\rm Tr}\Pi D+\Phi[D],
\eeq
where $\Phi$ is the sum of all the two-particle irreducible
(skeletons) diagrams. The self-energy
$\Pi$ is related to the propagator by Dyson's equation
\beq\label{sc1}
\,D^{-1}\,=\,D^{-1}_0\,+\,\Pi,
\eeq
and $D_0$ is the bare propagator. A remarkable feature of the
functional (\ref{functional}) is its stationarity property, i.e.,
${\delta {\cal F}[D] /\delta D}\,=\,0$ when 
\beq\label{sc2}
\Pi[D]\,=\,2\frac{\delta \Phi}{\delta D}\,.
\eeq
Together Eqs.~(\ref{sc1}) and (\ref{sc2}) form a self-consistent
set of equations which determine the full propagator. A major
observation by Baym
 is that self-consistent approximations can be
defined for any selection of skeleton diagrams in $\Phi$;
furthermore such ``$\Phi$-derivable'' approximations have the
property  to respect the  (global) symmetries of the
 hamiltonian 
\cite{Baym:1962}.  

The stationarity property of the free energy entails
important simplifications in the calculation of the entropy.
Indeed we have
\bea
{\cal S}\,=\,-\,{{\rm d}{\cal F}\over {\rm d}T}\,=\,
-\,{{\partial }{\cal F}\over {\partial T}}\bigg|_D
\eea
where the last step uses the fact that the temperature dependence
of the propagator can be ignored. Explicitly one gets:
\bea
{\cal S}=-\int\!\!\frac{d^4k}{(2\pi)^4}\frac{\partial N(\omega)}{\partial T}
{\rm Im}
\log D^{-1}(\omega,k) 
  +\int\!\!\frac{d^4k}{(2\pi)^4}\frac{\partial N(\omega)}{\partial T}
{\rm Im}\Pi(\omega,k)
{\rm Re} D(\omega,k)+{\cal S}'
\eea
where $N(\omega)=1/({\rm e}^{\beta \omega}-1)$, and  
$$
{\cal S}'\equiv -\frac{\partial (T\Phi)}{\partial T}\Big|_D+
\int\!\!\frac{d^4k}{(2\pi)^4}\frac{\partial N(\omega)}{\partial T} {\rm Re}\Pi\,
{\rm Im} D. 
$$
 A further simplification
occurs when one restricts the choice of skeletons to low order
ones  \cite{Riedel:1968,CP2,Vanderheyden:1998ph}. 
Thus, for the two-loop skeletons in QCD, 
$
{\cal S}'=0.
$

\section{The 2-loop entropy}
At two loop in the skeleton expansion, the entropy takes then the simple
form
\cite{Vanderheyden:1998ph,Blaizot:1999ip,Blaizot:1999ap,Blaizot:2000fc}:
\beq\label{entrop1}
S\,=\,-\int\frac{{\rm d}^4p}{(2\pi)^4}\,\frac{\partial 
N }{\partial T}\, \left\{{\rm
Im}\ln D^{-1}\,-\,{\rm Im}\Pi\,{\rm Re}D\right\}
\eeq
The simplifications discussed above have led to important cancellations
leaving for the entropy an expression which is effectively a one-loop
 expression, thus emphasizing the direct relation between the entropy and
the quasiparticle spectrum.  Residual interactions start contributing at
order 3-loop. This expression is also   manifestly ultraviolet-finite. 

Albeit simple, the 2-loop expression for the entropy is nevertheless
difficult to use as it stands. Recall that the propagator and self-energy in
this equation are solution of a self-consistent equation which is in general difficult to
renormalize and to solve.  Besides, in the case of gauge theories, the scheme does not
fully respect gauge symmetry. We have  therefore looked for additional approximations
which exploits the simplicity of the scheme (the one-loop feature of the entropy formula)
and at the same time allows for  practical calculations. The main idea is to find gauge
invariant approximations for the self-energy by exploiting  kinematical approximations 
such as those which lead in particular to the hard thermal loops. The entropy
is then calculated exactly with no further approximation. 
One constraint that we impose on our approximate $\Pi$ is that, once inserted in the
formula (\ref{entrop1}),  the resulting value of $S$ is perturbatively correct up
to order
$g^3$ (which is the maximum perturbative accuracy that one can achieve with
2-loop skeletons). 

 In the regime where the loop momenta
are soft, we can use as an approximation for $\Pi$ the corresponding hard
thermal loop (HTL), that is 
$$\omega,\,p\sim gT\qquad\Pi_{soft}\,\approx\,\Pi_{HTL}$$
For hard momenta on the other hand, the hard thermal loop is  
no longer valid:  corrections to hard particle dispersion relations due to their coupling
with soft modes need to be taken into account.  This can be
estimated using usual HTL perturbation theory. A simplification occurs however because  we
need such corrections only near the light cone: 
$$\omega,\,p\sim T\qquad\Pi_{hard}(\omega^2\sim p^2)$$

In fact we shall find convenient to define two successive approximations.
In the first one, called the HTL approximation, we use 
$\Pi=\Pi_{HTL}$ at 
 all  
momenta. This is a fine approximation for soft momenta, but it is also a
decent approximation at hard momenta because the self-energy is needed only
in the vicinity of the light cone where the hard thermal loop approximation
 coincides with the full one-loop result. We call
${\cal S}_{HTL}$ the resulting expression for the entropy. This expression
fully accounts for the perturbative contributions of order
$g^2$ but for only $1/4$ of that of order $g^3$. In the  next-to-leading
approximation one uses 
$\Pi_{soft}=\Pi_{HTL}$ and  $\Pi_{hard}=\Pi_{HTL}+\delta \Pi$, where $\delta \Pi$
contains the coupling to the soft modes (I should emphasize  
at this point that the construction of our next-to-leading approximation involves several 
subtle steps which I am skipping here but which are detailed by A. Rebhan in his contribution \cite{RebhanHere}). We call
${\cal S}_{NLA}$ the corresponding entropy; it takes fully into account all perturbative
contributions of order
$g^2$ and
$g^3$.

As an illustration of the quality of the results which we can obtain, I
present in Fig.~3 the entropy of pure SU(3)
 gauge theory normalized to the ideal gas entropy. The
agreement at large $T$ ($T\simge 2.5 T_c$) is quite good. Note
also that in going from one level of approximation to the next
(i.e. from ${\cal S}_{HTL}$ to ${\cal S}_{NLA}$), the changes
are moderate, in contrast to what happens in ordinary
perturbation theory. This reflects the stability of the present
scheme. It also points to the fact that the contribution of the
soft collective modes is indeed small: the fact that they give
a  seemingly large contribution at order $g^3$ in perturbation
theory is just an artifact of the truncation of a resummed
expression at finite order.

\begin{figure}[h]\label{fig:pl1a}
\includegraphics[bb = 70 180 540 540,width=10.5truecm]{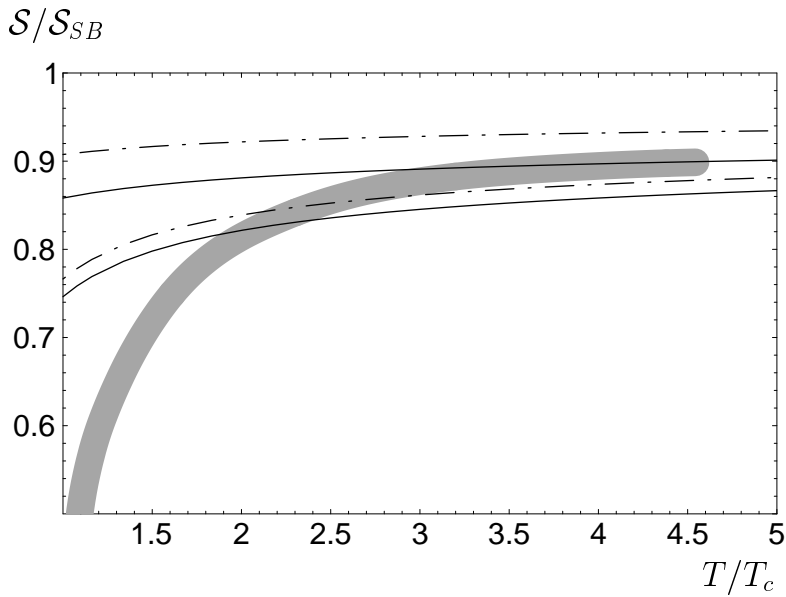}
\caption{The entropy of pure SU(3)
  gauge theory normalized to the ideal gas entropy
${\cal S}_{SB}$.
  Full lines: ${\cal S}_{HTL}$.
Dashed-dotted lines: ${\cal S}_{NLA}$.
 For each approximation, the two lines correspond to the
choices $\bar\mu=\pi T$ and $\bar\mu=4\pi T$ of the running
coupling constant
$\alpha_s(\bar\mu)$ in the $\overline{{\rm MS}}$
renormalization scheme.
The dark grey band represents the  lattice results of Ref.~[2].
}
\end{figure}

\section{CONCLUSIONS}

Our results confirm that for temperatures larger than about
3$T_c$, the thermodynamical functions of the quark-gluon
plasma, in particular its entropy, can indeed be interpreted as
those of weakly interacting quasiparticles. The dominant effect
of the interactions is to modify the spectrum of these
quasiparticles. As the coupling grows the quasiparticle
properties are non perturbatively renormalized but their mutual
interactions remain  weak. 

Thermodynamic functions are dominated by hard degrees of
freedom. While long range correlations may survive in the
quark-gluon plasma at very large temperature, the present
picture suggests that such correlations do not contribute
much to the thermodynamics. We find explicitly that the soft 
contributions are indeed small. We argued in particular that the
contribution of collective modes is artificially amplified when
the coupling is not too small by truncating this contribution
at order
$g^3$.
 
The approach that we have developed relies on a hierarchy
of scales that emerges when the coupling is small. It is
an assumption that the structure identified at weak coupling
survives when the coupling  grows, e.g.  as we lower the
temperature. The fact that we find approximations where such an
extrapolation works supports  the validity of
this assumption.  But of course a complete check of the method can
only be done by comparing with ``exact'' results, such as those
provided by lattice techniques. Much can be learned also from detailed 
comparisons with the other approaches discussed earlier, namely
hard thermal loop perturbation theory or dimensional
reduction. There is finally much to do also within the present theory
itself,  to push for instance non perturbative renormalization
techniques (interesting  progress in this direction has
been reported recently \cite{vanHees:2001ik}; see also
\cite{Braaten:2001vr}). 

Further tests involve the calculations of different 
observables. An example, advocated by Gavai
\cite{GavaiHere} at this workshop, is that of quark
susceptibilities. This is a calculation that we have recently
taken up, with interesting results \cite{Blaizot:2001vr}. 
The calculation of quark susceptibilities represents a small
incursion into the physics of finite chemical potentials which
is of course readily available in our approach \cite{RebhanHere}.

\section*{Acknowledgements}

The work presented here has been carried out in a most
enjoyable collaboration
with E. Iancu  and A. Rebhan.
I would also like to thank  Frithjof Karsch and Helmut Satz,
for their invitation to this very stimulating meeting.


\bibliography{tft,tftpr,qft,ar,books}      
\bibliographystyle{elsart-numwot}

\end{document}